\documentclass[
    ,final            % use final for the camera ready runs
  ]
  {aipproc}

\layoutstyle{8x11double}
\usepackage{amssymb}
\usepackage{wasysym}
\begin{document}

\title{Glasses of dynamically asymmetric binary colloidal mixtures: Quiescent properties and dynamics under shear}


\author{Tatjana Sentjabrskaja}{
  address={Condensed Matter Physics Laboratory, 
  Heinrich Heine University D\"usseldorf, 40225 Germany}
}

\author{Donald Guu}{
  address={ICS-3, Institut Weiche Materie, Forschungszentrum Jülich, 52425 Jülich, Germany.} % additional visiting address
}

\author{M Paul Lettinga}{
  address={ICS-3, Institut Weiche Materie, Forschungszentrum Jülich, 52425 Jülich, Germany.} % additional visiting address
}

\author{Stefan U Egelhaaf}{
  address={Condensed Matter Physics Laboratory,
  Heinrich Heine University D\"usseldorf, 40225 Germany} % additional visiting address
}

\author{Marco Laurati}{
  address={Condensed Matter Physics Laboratory,
  Heinrich Heine University D\"usseldorf, 40225 Germany}
}

\begin{abstract}
 We investigate mixing effects on the glass state of binary colloidal hard-sphere-like mixtures with large size asymmetry, at a constant volume fraction $\phi = 0.61$. The structure, dynamics and viscoelastic response as a function of mixing ratio reflect a transition between caging by one or the other component. The strongest effect of mixing is observed in systems dominated by caging of the large component. The possibility to pack a large number of small spheres in the free volume left by the large ones induces a pronounced deformation of the cage of the large spheres, which become increasingly delocalised. This results in faster dynamics and a strong reduction of the elastic modulus. When the relative volume fraction of small spheres exceeds that of large spheres, the small particles start to form their own cages, slowing down the dynamics and increasing the elastic modulus of the system. The large spheres become the minority and act as an impurity in the ordering beyond the first neighbour shell, i.e. the cage, and do not directly affect the particle organisation on the cage level. In such a system, when shear at constant rate is applied, melting of the glass is observed due to facilitated out-of-cage diffusion which is associated with structural anisotropy induced by shear.  
 \end{abstract}

\maketitle

%%%%%%%%%%%%%%%%%%%%%%%%%%%%%%%%%%%%%%%%%%%%
%% MAINMATTER
%%%%%%%%%%%%%%%%%%%%%%%%%%%%%%%%%%%%%%%%%%%%

\section{Introduction}

Many different systems, among them polymers, metals and colloids, can form thermodynamically equilibrated states, but also non-equilibrium, metastable states, including amorphous solid materials called glasses \cite{dhont_book,pusey:leshouches}. 
The glass transition is generally associated with a dramatic slowing down of the particle dynamics which is driven by changes in thermal energy or crowding.\\ 
One of the simplest model systems to study crowding induced glass formation are suspensions of colloidal hard spheres. 
By increasing the particle volume fraction $\phi$ formation of a glass state above $\phi = \phi_g$ prevents crystallisation, if the system has a sufficiently broad distribution of sizes. 
The formation of the glass state is explained in terms of the cage effect: At $\phi > \phi_g$ each particle is trapped in the cage of its neighbours resulting in dynamical arrest, i.e. the absence of long distance diffusion over a large window of times \cite{pusey_vm_prl87,vm_pre98}. 
Dynamical arrest and formation of a solid state above $\phi \geq \phi_g$ are also manifested in the viscoelastic properties as a sudden increase of the viscosity \cite{larson99} and the appearance of a Maxwell plateau modulus in the linear response \cite{mason_prl95}. 

The addition of a second component with a significantly different mean size compared to the first component, leads to an even richer scenario. Depending on the total volume fraction of the system and the mixing ratio of the two species, mode coupling theory (MCT) predicts the existence of different glass states \cite{Voigtmann.2011b}. 
When the size-ratio $\delta = R_s/R_l = d_s/d_l$, where $R_s$, $d_s$ and $R_l$, $d_l$ are the radii and diameters of the small and large components respectively, becomes about 0.2 and smaller, four different glass states are expected \cite{Voigtmann.2011b}: In the first state both components are caged; in the second state dynamical arrest of the large component is driven by depletion attraction induced by the small species; in the third state the large component is arrested through caging, while the small component is mobile; finally the small particles can be caged, while the large particles are not caged, but only localised by the surrounding dense matrix of small particles.
Despite the rich behaviour predicted by theory, the glass state of colloidal binary mixtures at such large size disparities is hardly studied experimentally \cite{Imhof.1995}. In \cite{Imhof.1995} the formation of a glass despite the mobility of the small component is reported. A similar glass state has also been found in simulations of soft sphere mixtures  \cite{Moreno.2006,Moreno.2006b}. 

In order to extend these studies and to explore the formation of different glasses, we performed experiments to determine the microscopic structure, dynamics and viscoelastic response of colloidal hard-sphere mixtures of large size disparity ($\delta = 0.2$) and constant total volume fraction $\phi \simeq 0.61$. 
We vary the relative volume fraction of the small component, $x_s = \phi_s/\phi$, to explore the effect of mixing on the glass state. We find that the composition of the mixture strongly affects the dynamics and elastic modulus of the system, in particular in mixtures containing a smaller volume fraction of small spheres, $x_s < 0.5$.  
In addition, we compare the dynamics of a sample under shear to its quiescent state, showing that the driving introduced by shear leads to an acceleration of the non-affine particle motions, inducing glass melting. 
 A discussion of the non-linear rheology of these mixtures and comparison to predictions of mode-coupling theory are reported in separate publications \cite{MarcoNonlin,thomas_draft_lin}.

\section{Methods}

\subsection{Samples}

Suspensions of poly-methylmethacrylate (PMMA) particles sterically stabilized with a layer of polyhydroxystearic acid (PHS) were prepared in a solvent mixture of cycloheptyl bromide (CHB) 
and cis-decalin, closely matching the density and refractive index of the colloids. 
In the CHB/decalin solvent mixture, the spheres acquire a small charge which is screened by adding 4 mM tetrabutylammoniumchloride \cite{yethiraj03}. 
This system shows almost hard-sphere behaviour, with the volume fraction $\phi=(4\pi/3)nR^3$ being the only thermodynamic control parameter, with $n$ the number density of particles and $R$ the sphere radius.
Binary colloidal mixtures with $\delta\simeq0.2$, fixed total volume fraction $\phi \simeq 0.61$ and different mixing ratios were prepared starting from one component stock suspensions. The stock suspensions were obtained by diluting a sediment of large particles of mean size $d_l = 1.76\pm 0.02$ $\mu$m (relative polydispersity $\sigma$ = 0.057), or small particles of mean size $d_s=0.350\pm 0.004$ $\mu$m ($\sigma = 0.150$). The large particles were fluorescently labeled with nitrobenzoxadiazole (NBD).
For the two one-component colloidal stock suspensions, the values of the radius and polydispersity were determined from the angular dependence of the scattered intensity and diffusion coefficient obtained by means of static and dynamic light scattering, respectively, on a very dilute colloidal suspension ($\phi\simeq 10^{-4}$). 
The volume fraction of the sediment of large spheres was experimentally determined as follows: A first guess for the volume fraction $\phi_{RCP}$ of the sediment was obtained using simulation results \cite{schaertl94}. The sediment was then diluted to a nominal volume fraction $\phi\simeq 0.4$ and observed using confocal microscopy. 
The imaged volume was partitioned into Voron{\"o}i cells and the mean size of the Voron{\"o}i volume per particle calculated. 
The ratio of the particle volume to the mean Voron{\"o}i volume serves as an estimate of the volume fraction of the sample. 
This was found to be $\phi = 0.43$ which corresponds to $\phi_{RCP}^l = 0.68$. 
The small spheres were too small to be imaged. Therefore their volume fraction was adjusted to give an equivalent rheological response to the large spheres. 
For ideal hard spheres, the energy density scales as
$nk_BT$, so that the shear moduli must be equal in these units.
The volume fraction of the one-component small particles suspension was adjusted to obtain the same normalised shear moduli as for the one-component large particles. 
Although their
linear viscoelasticities are thus within experimental resolution the same, their volume fractions could be slightly different,
since the samples have different polydispersities.
Accordingly, for intermediate $x_s$, shear moduli are reported in reduced units
of the energy density.

\subsection{Confocal microscopy}
\subsubsection{Quiescent State}
\begin{figure}
  \includegraphics[width=7cm]{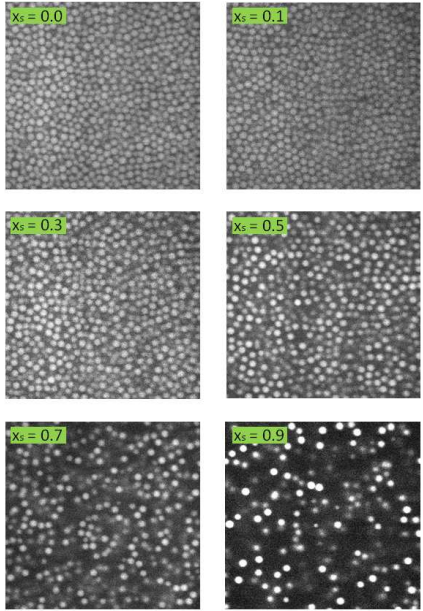}
  \vspace{-3cm}
	\caption{Typical confocal microscopy images of quiescent samples showing the large fluorescently-labelled particles only. The total volume fraction of the samples is $\phi = 0.61$ and the relative volume fraction of small particles $x_{\mathrm{s}}$ are as indicated.}
  \label{conf_images}
\end{figure}

Confocal microscopy experiments on quiescent samples were performed using a VT-Eye confocal unit (Visitech International), mounted on a Nikon Ti-U inverted microscope with a 100x Nikon Plan-Apo VC oil-immersion objective, and a laser with $\lambda = 488$~nm. Samples were contained in vials where the bottom was cut and replaced by a coverslip to allow for imaging \cite{jenkins08}. Stacks of images of 512$\times$512 pixels, corresponding to an x-y plane size of approx.\ $50\times50$ $\mu$m$^2$ were acquired. Each stack was composed of 101 images obtained every 0.2~$\mu$m in z-direction, leading to an imaged volume of approximately\ $50\times50\times20$ $\mu$m$^3$ per stack. The time needed to acquire one stack was approximately 3.8~s. Stacks were acquired at a depth of approx.\ 30$ \mu$m from the coverslip.
Typically for each sample 7 different volumes were imaged for 1200s during which 300 stacks were collected for each volume to follow the dynamics of the samples.  
The stacks were analysed using standard routines \cite{crocker96} to extract particle coordinates and trajectories.
Figure \ref{conf_images} shows typical two-dimensional images corresponding to a plane in a stack, acquired for samples with different mixing ratios $x_s$.

\subsubsection{Under Shear}
Under shear, samples were imaged using a custom-built rotational shear cell (a modified version of the model described in \cite{derks04}), mounted on a Zeiss Axiovert M200 microscope with a 63x Zeiss Plan Neo Fluar water-glycerol immersion objective and equipped with a VT-Infinity confocal unit (Visitech International).
A glass coverslip serves as bottom plate of the cell to allow for imaging with high numerical aperture objectives. The glass surface was covered with polydisperse colloidal hard spheres with a size comparable to the large spheres to minimise the effects of wall slip \cite{ballesta_2012}.
The top of the cell is formed by a metal cone with 14~mm diameter and 2$^{\circ}$ cone angle. Images are acquired at a radial distance of 7~mm from the center. 
The plate and cone rotate in opposite directions, giving rise to a zero-velocity plane in the sample, the depth of which can be adjusted through the relative speed of the cone and plate. 
Images were acquired with an Andor iXon 897 EMCCD camera for 300s, at an average rate of 10 frames per second. 
Solvent evaporation was minimised using a solvent sealing at the top of the cell.

\subsection{Rheology}
Rheology measurements were performed with a AR2000ex stress-controlled rheometer, using a cone-plate geometry with 20~mm diameter, 2$^{\circ}$ cone angle and 0.054~mm gap. 
A solvent trap was used to minimise solvent evaporation during the measurements. 
The temperature was set to 20 $^{\circ}$C and controlled within $\pm$0.1 $^{\circ}$C via a standard Peltier plate. 
The effects of sample loading and aging were minimized by performing a standard rejuvenation procedure before each test: directly after loading, we performed a dynamic strain sweep, i.e. applied oscillatory shear to the samples with a frequency $\omega$ = 1 rad/s and an increasing strain amplitude until the sample was flowing.  
Before each measurement, flow of the sample was induced applying oscillatory shear at strain $\gamma$ = 300\%. Shear was applied for the time needed to achieve a steady-state response, i.e. the storage modulus G$^\prime$ and the loss modulus G$^{\prime\prime}$ become time-independent, typically 200~s. Successively, the linear viscoelastic moduli were measured at $0.1$\% $\leq\gamma\leq 0.5$\% (depending on sample) as a function of time to monitor reformation of structure, until the moduli reached a time-independent value, typically after 100~s to 900~s (depending on sample). After this, the experiment was started immediately.

\section{Results and Discussion}
\subsection{Quiescent Structure}
\begin{figure}
  \includegraphics[scale=0.2]{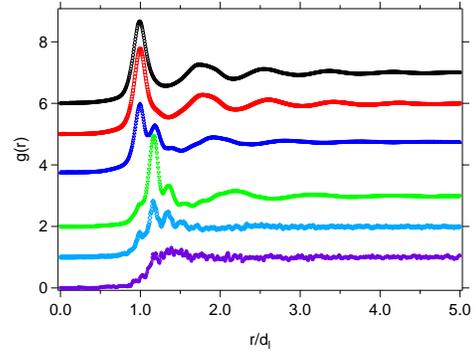}
  \caption{Radial distribution function $g(r)$ of the large spheres, determined using confocal microscopy. The total volume fraction of the samples is $\phi = 0.61$ and the relative volume fractions of small particles are $x_S$ = 0, 0.1, 0.3, 0.5, 0.7, 0.9 (top to bottom). Data are shifted along the vertical axis for clarity.}
  \label{gr}
\end{figure}

To understand mixing-induced changes on the cage structure of a one-component glass, we used confocal microscopy to determine the radial distribution functions $g(r)$ of large spheres, the only species which is fluorescently labeled and therefore visible (Figs.~\ref{conf_images} and \ref{gr}).

%\paragraph{<A subsubsubsection>}
%%%%%%%%%%%%%%%%%%%%%%%%%%%%%%%%%%%%%%%%%%%%
%% Sample figure:
%%
%% The option [height=...] scales the picture to the given height,
%% without it it would be printed at its nominal size
%%%%%%%%%%%%%%%%%%%%%%%%%%%%%%%%%%%%%%%%%%%%

The $g(r)$ 
for $x_s=0$ is typical of a glass-forming one-component suspension
with size polydispersity. It shows
a pronounced correlation peak at $r\approx d_l$, corresponding to the highest probability of finding particles in the
first-neighbour shell, and additional peaks at larger $r$ related to particles in the successive neighbour shells. 
For a small volume fraction of small particles ($x_s=0.1$, $\phi_l = (1-x_s)\phi = 0.549$) these features remain, but in addition a small shoulder to the right of the first maximum is observed. 
This indicates a perturbation of the cage formed by the large spheres. 
When increasing $x_s$ to 0.3 ($\phi_l = 0.427$), the
height of the first-neighbour peak decreases, which indicates dilution (also evident in Fig. \ref{conf_images}), and that some particles formerly constituting the cage are located at larger distances. 
These particles are found at distances $d_l+d_s$ (where the shoulder was observed at $x_s = 0.1$) and $d_l+2d_s$, as seen from the corresponding peaks in g(r). 
This implies that small particles are located in between large particles and hence loosen the cage structure. 
In line with this observation the layering of large spheres only extends to the third neighbour shell.  
At $x_s = 0.5$ ($\phi_l = 0.305$) particles are mostly located at distance $d_l+d_s$ and also the probability of finding particles at $d_l+2d_s$ is increased. 
Moreover, additional peaks at $d_l+n\,d_s$ are visible. 
This indicates that at $x_s = 0.5$ a first neighbour shell of large spheres does no longer surround large particles (Fig. \ref{conf_images}), and a transition to a cage of small spheres takes place.
This is consistent with the following geometrical argument: Each small sphere of radius $R_s$ projects on a sphere of radius $R_l=R_s/\delta$ an angle $\theta = 2\arcsin(1/(1+1/\delta))$. The maximum packing of small spheres having the centers separated by this angular distance, i.e. covering the surface area of a large sphere, can be calculated as $N = 120$ \cite{sloane:book}.
At $x_s = 0.5$ the number fraction of small particles for each big particle is $\xi_s/\xi_l=x_s/\delta^3(1-x_s)=125$, i.e. on average each large particle is covered by small particles for $x_s = 0.5$ and hence the first neighbour shell and cage of large particles disappear.
At $x_s = 0.7$ ($\phi_l = 0.183$) correlations at distances $d_l+n\,d_s$ dominate and layering beyond the second neighbour shell vanishes due to the pronounced dilution of the large spheres (Fig. \ref{conf_images}).
Correlations are further reduced at $x_s = 0.9$ ($\phi_l = 0.061$) due to the increased dilution of the large spheres (Fig. \ref{conf_images}).
From the evolution of the radial distribution function with increasing $x_s$ one can therefore conclude that the small spheres occupy an increasingly larger fraction of the free volume in between the large spheres, inducing a distortion of the cage of large spheres until a transition to a system dominated by the cage of small spheres is observed.

\subsection{Quiescent Dynamics}

\begin{figure}
  \includegraphics[scale=0.2]{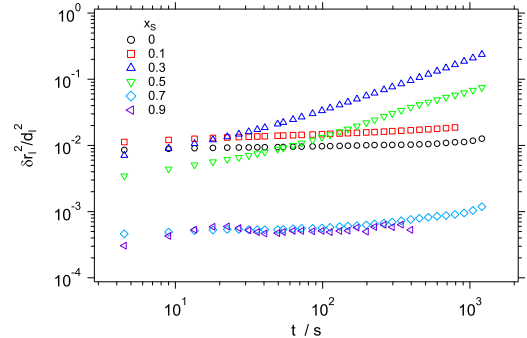}
  \caption{Mean squared displacements $\delta r_l^2$ of the large spheres, normalised by their diameter squared ($d_l^2$), determined using confocal microscopy. The total volume fraction of the samples is $\phi = 0.61$ and the relative volume fractions of small particles are $x_S = 0.0$ ($\circ$), 0.1 ($\Box$), 0.3 ($\triangle$), 0.5 ($\triangledown$), 0.7 ($\Diamond$), 0.9 ($\triangleleft$).}
  \label{msd}
\end{figure}

To explore the effect of the structural changes discussed in the previous section on the microscopic dynamics, we investigate the mean squared displacements (MSDs) of the large particles, $\delta r_l^2$, as a function of $x_s$ (Figure \ref{msd}). 
The system of only large spheres ($x_s = 0.0$) presents an MSD which, within the accessible time range, shows no long-time diffusion, i.e. glassy dynamics. 
Moreover, the plateau of the MSD corresponds to a localisation of the particles on distances of the order $\nu_l (x_S = 0)\approx 0.1 d_l$, which is typical for a cage in a one-component glass.
The time-dependence of the dynamics is similar for $x_s = 0.1$, but the localisation length is slightly larger. This reflects the small perturbation of the cage structure (Figure \ref{gr}).
For $x_s = 0.3$ a significant acceleration of the dynamics is observed, for times $t > 10$~s the particles are no longer localised and the MSD increases sub-linearly with $t$. 
It is expected that diffusive dynamics is established beyond the accessible time scale.
The acceleration of the dynamics is related to the considerable distortion of the cage of large spheres induced by the presence of the small spheres, which increase the mobility.
A comparable time-dependence of the MSD is obtained at $x_s = 0.5$, but the displacements are smaller, indicating a stronger localisation. A stronger localisation can be associated with the transition to the cage structure of small spheres, as also evidenced in the $g(r)$. 
Note that the caging of the small spheres is apparently incomplete yet, and therefore the large particles are not localised.
For the two largest values of $x_s$ the large spheres are localised by a cage of small spheres. Accordingly their MSD again show no diffusion.
The localisation length is of the order of $\nu_l(x_s \geq 0.7,0.9)\approx 0.02 d_l$, i.e. about $\delta\nu_l (x_S = 0)$, which indicates that the particles are indeed localised on the length scale of the cage of small particles. Note that the plateau values of the MSDs of these samples approach the resolution limit of the setup. 

\subsection{Linear Viscoelastic Moduli}
\begin{figure}
  \includegraphics[scale=0.2]{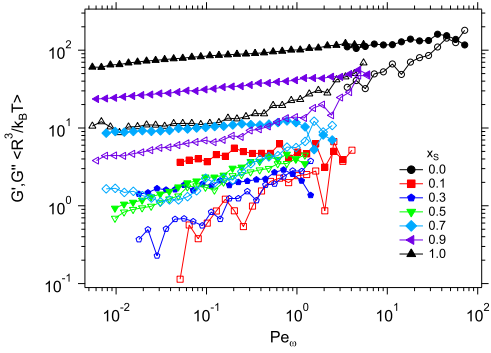}
  \caption{Storage modulus G$^{\prime}$ (full symbols) and loss modulus G$^{\prime\prime}$ (open symbols), in units proportional to the energy density ($k_BT/\langle R^3\rangle$), as a function of oscillatory Peclet number $Pe_{\omega}$, for samples with $\phi = 0.61$ and relative volume fractions of small particles $x_S = 0.0$ ($\circ$), 0.1 ($\Box$), 0.3 ($\pentagon$), 0.5 ($\triangledown$), 0.7 ($\Diamond$), 0.9 ($\triangleleft$), 1.0 ($\triangle$).}
  \label{gw}
\end{figure}   
In order to establish a link between the microscopic structure and dynamics of the samples and their mechanical response, we measured the frequency dependent linear viscoelastic moduli of the mixtures (Figure \ref{gw}). The moduli are reported in units of energy density $\langle nk_BT\rangle\sim k_BT/\langle R^3\rangle$, with $n$ particle density and:  
\begin{equation}
\frac{1}{\langle R^3\rangle}=\frac{1}{R_L^3}\big[x_S\big(\frac{1}{\delta^3}-1\big)+1\big]
\end{equation}
This representation removes the trivial effect of different average particle sizes for different values of $x_\mathrm{s}$ on the absolute values of the shear moduli. The data are shown as a function of the oscillatory Peclet number $Pe_{\omega}=\tau_B/\tau_{\omega}=(6\pi\eta\omega\langle R^3\rangle)/ k_BT$ which represents the ratio between the period of oscillation, $\tau_{\omega}=1/\omega$, and the Brownian time, $\tau_B=\langle R^2/D_0\rangle$, where $D_0$ is the free diffusion coefficient.\\
At large values of $Pe_{\omega}$,  for all samples $G^{\prime\prime}$ is larger than $G^{\prime}$. This response can be associated to the in-cage dynamics, i.e. the short time diffusion of a particle in its cage. In contrast at smaller frequencies, i.e. longer times, the structural relaxation associated with long-time diffusion allows us to distinguish the response of a glass from that of a fluid.\\
The one-component systems ($x_s = 0.0,1.0$) show the response of a glass. The storage modulus $G^{\prime}$ is larger than the loss modulus $G^{\prime\prime}$ 
 and no crossing of the two moduli can be observed at low $Pe_{\omega}$, indicating that no structural relaxation is observed in the accessible frequency window. 
At $x_s = 0.1$ the $Pe_{\omega}$ (frequency) dependence of the viscoelastic moduli is similar to that of the one-component systems, i.e. still characteristic of a glass, but the viscoelastic moduli are reduced by more than an order of magnitude, despite the only limited structural deformation of the cage induced by the presence of the small spheres. 
This is however consistent with the larger localisation length observed in the dynamics, which indicates a looser cage structure.  
At $x_s = 0.3$ the reduction of the shear moduli is even more pronounced than at $x_s = 0.1$. 
Moreover $G^{\prime}$ and $G^{\prime\prime}$ become similar, indicating a weaker solid-like response. 
This is consistent with the large structural distortion of the cage manifested in the radial distribution function (Figure \ref{gr}) and with the faster dynamics (Figure \ref{msd}).
Compared to $x_s = 0.3$, at $x_s = 0.5$ the moduli at large $Pe_{\omega}$ are bigger, but smaller at low $Pe_{\omega}$, which is due to the stronger frequency dependence of the shear moduli. 
In addition, $G^{\prime}$ and $G^{\prime\prime}$ have almost identical values. 
Such a response is similar to that observed for depletion driven colloidal gels in the vicinity of the gelation boundary \cite{chambon86,winter86,laurati_jcp09}.
Further increasing $x_s$ to 0.7 and 0.9, the moduli become larger and for $x_s = 0.9$ approach the values of the one-component systems. 
This is consistent with a transition to a system dominated by cages formed by the small particles and with structure and dynamics of these samples. 
Note that the residual distortion induced by the large spheres leads to a reduction of the overall elastic response of these samples.\\
\begin{figure}
  \includegraphics[scale=0.4]{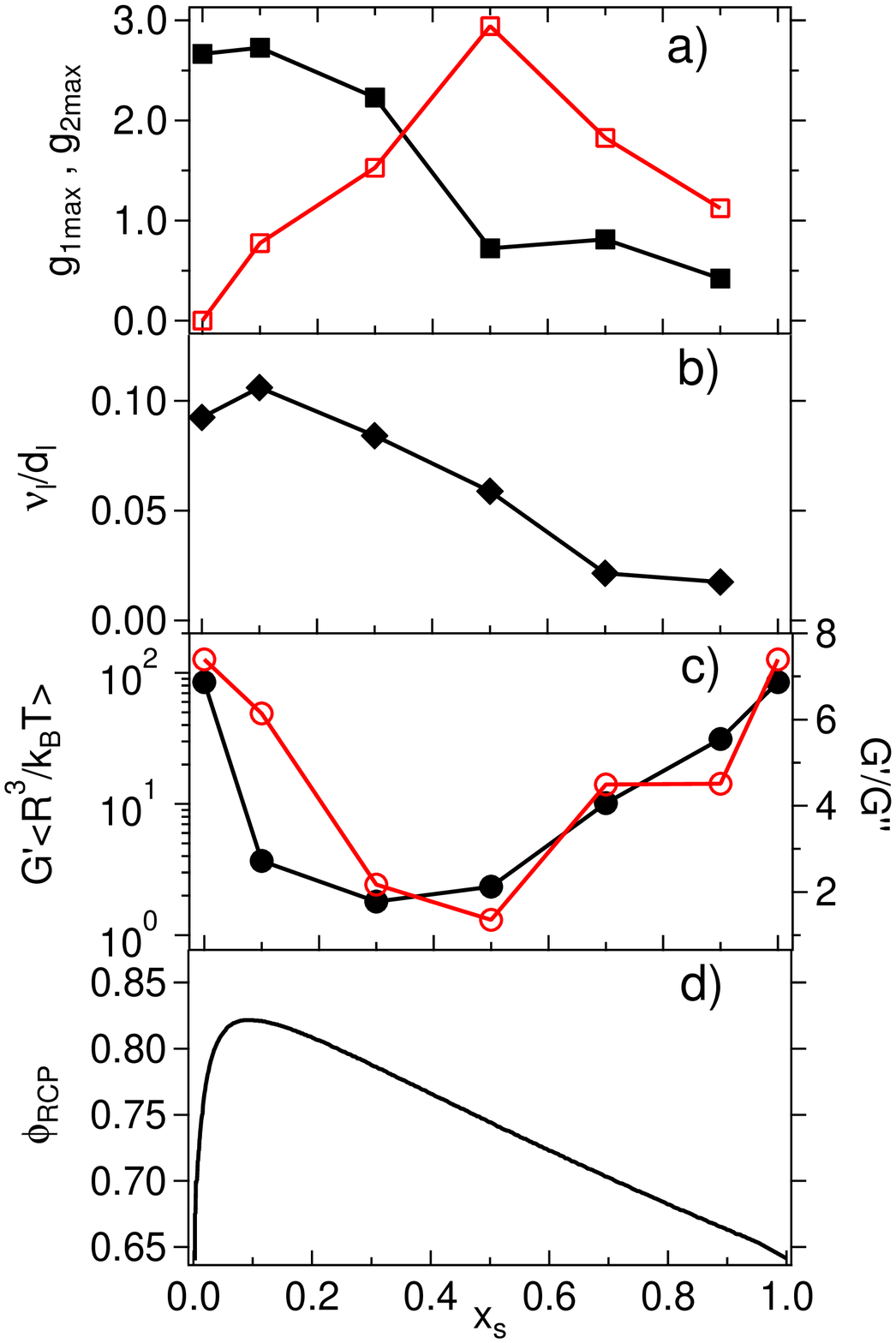}
  \caption{(a) Height of the first, $g_\mathrm{1max}$ (black, full symbols), and second, $g_\mathrm{2max}=g(d_\mathrm{l}+d_\mathrm{s})$ (red, open symbols), peaks of $g(r)$ (Figure \ref{gr}), (b) Localisation length $\nu_l$ in units of $d_l$, estimated from MSDs at $t=4.5\,s$ (Figure \ref{msd})  (c) Storage modulus $G^{\prime}(Pe_{\omega} = 0.1)$ (black, full symbols, left axis) and ratio $G^{\prime}/G^{\prime\prime}$ at the same $Pe_{\omega}$ (red, open symbols, right axis), in units of energy density $k_BT/\langle R^3\rangle$) and (d) theoretical prediction \cite{biazzo:2009} for changes in $\phi_{RCP}$ at $\delta = 0.175$ as a function of $x_s$.}
  \label{resume_plot}
\end{figure}        
The trends discussed above are summarized by plotting $G^{\prime}$ and the ratio $G^{\prime}/G^{\prime\prime}$ as a function of $x_s$, at a fixed value of $Pe_{\omega} = 0.1$ (Figure \ref{resume_plot}c). 
The ratio $G^{\prime}/G^{\prime\prime}$ attains the smallest value at $x_s = 0.5$, which could reflect a transition from a system dominated by cages formed by large spheres to a system dominated by cages formed by small spheres. 
This interpretation is supported by the trends of the heights of the first and second peaks of $g(r)$ (Figure \ref{resume_plot}a): Between $x_s = 0.3$ and $x_s = 0.5$ the first peak strongly drops and then remains nearly constant for larger $x_s$, indicating the disappearance of the first neighbor shell of large particles, i.e the large spheres cage. The second peak reaches its maximum at $x_s = 0.5$, corresponding to formation on average of a shell of small particles around each large particle, and then decreases for larger values of $x_s$, due to the further intercalation of small spheres in between two large spheres, which leads to caging of the small spheres. 
In contrast, the elasticity of the samples, represented by $G^{\prime}$, reaches a minimum at $x_s = 0.3$. 
This could be explained by the larger localisation length of the large spheres at $x_s = 0.3$ (Figure \ref{resume_plot}b). 
Furthermore, changes in $G^{\prime}$ are considerably larger in systems with a larger volume fraction of large spheres. 
This can be rationalized by considering the effects of the inclusion of the second component on the structure of the system in the two cases: In systems at small $x_s$, the small spheres can be packed in the free volume in between the large spheres, including the free volume within the cages. 
This deforms the cage and shifts random close packing (Figure \ref{resume_plot}d, data for $\delta = 0.175$ taken from \cite{biazzo:2009}). 
On the other hand, addition of large spheres to a system of small spheres only affects the order beyond the first shell, i.e. beyond the cage, since the large spheres cannot fill the space in between the small spheres. 
This results in structural heterogeneity rather than cage deformation, and in a small shift of random close packing (Figure \ref{resume_plot}d, $x_s > 0.5$).
 
\subsection{Dynamics under shear}

\begin{figure}
%  \vspace{-3cm}
  \includegraphics[scale=0.3]{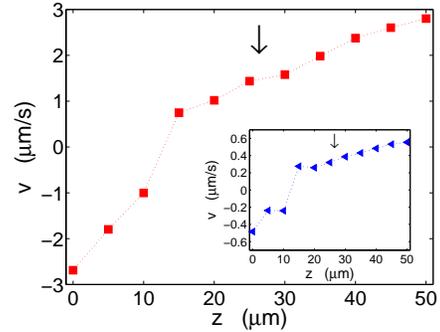}
 % \vspace{-3cm}
  \caption{Average velocity of the large spheres as a function of position $z$ in the gap, for sample with a relative volume fraction of small spheres $x_s = 0.9$ and $Pe_{\dot{\gamma}}^l = 4.2\times 10^{-1}$ (main plot) and $6.7\times 10^{-2}$ (inset). Arrows indicate the location of the plane where the dynamics were measured.}
   \label{vel_profiles}
\end{figure}

\begin{figure}
  \includegraphics[scale=0.2]{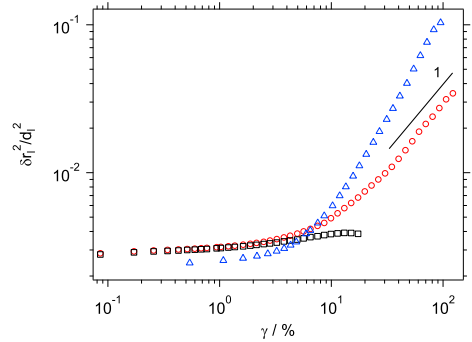}
  \caption{Mean squared displacements $\delta r_l^2$ of the large spheres as a function of strain $\gamma=\dot{\gamma} t$, in units of the squared large spheres diameter $d_l^2$, determined by confocal microscopy for sample with a relative volume fraction of small spheres $x_s = 0.9$, in the quiescent state ($\Box$) and  in the steady state of shear at 
  $Pe_{\dot{\gamma}}^l = 6.7\times 10^{-2}$ ($\circ$) and $4.2\times 10^{-1}$ ($\triangle$). The MSDs under shear only contain non-affine particle motions. The strain axis for the quiescent state was calculated using the lower shear rate $\dot{\gamma} \simeq 0.06$~s$^{-1}$.}
  \label{msd_shear}
\end{figure}

We investigated the effect of shear on the motions of large particles in a sample with  a major relative volume fraction of small spheres ($x_s = 0.9$). The mean squared displacements $\delta r_l^2$ of large particles were determined in the quiescent and steady state of shear, for two different shear rates $\dot{\gamma}$.  
For the applied shear rates, the time scale introduced by shear, $1/\dot{\gamma}$, is  considerably longer than the Brownian time $\tau_B$ associated with the short-time diffusion of both large and small spheres. 
This is quantified through the Peclet number $Pe_{\dot{\gamma}} = (6\pi\eta\dot{\gamma}\langle R^3\rangle)/k_BT$.
Both time scales are smaller than the structural relaxation time of the system, which diverges, although activated processes typically lead to diffusion at long times \cite{brambilla09}.\\
The velocity profiles obtained in the steady state of shear are shown in figure \ref{vel_profiles}. 
They were obtained by determining the velocity of the particles from their trajectories. 
The zero-velocity plane is located at about 15\,$\mu$m into the sample for both shear rates. 
The velocity profiles show a larger velocity gradient below the zero-velocity plane than above. Within each band though the velocity profile is linear, indicating laminar flow. 
For the higher shear rate the slower band corresponds to $\dot{\gamma} \simeq 0.06$~s$^{-1}$, while the faster band to $\dot{\gamma} \simeq 0.25$~s$^{-1}$. 
The weighted average $\dot{\gamma} \simeq 0.117$~s$^{-1}$ agrees within uncertainties with the expected value of $\dot{\gamma} \simeq 0.120$~s$^{-1}$.
Similarly for the slower shear rate the slower band corresponds to $\dot{\gamma} \simeq 0.0095$~s$^{-1}$ and the faster one to $\dot{\gamma} \simeq 0.041$~s$^{-1}$, with an average of 0.017~s$^{-1}$.
The formation of the two bands might be due to the different roughness of the particles coated surface of the bottom plate and the smooth metallic surface of the cone. 
The dynamics under shear were determined in a velocity-vorticity plane within the slower bands for both shear rates, corresponding for the large spheres to $Pe_{\dot{\gamma}}^l = 4.2\times 10^{-1}$ and $6.7\times 10^{-2}$, and at about $z = 27\,\mu$m in the sample (arrows in Figure \ref{vel_profiles}).\\    
The results of measurements with the shear cell setup are reported in Figure \ref{msd_shear} as MSDs vs. strain $\gamma = \dot{\gamma}t$. The strain axis for the quiescent MSD was obtained using the faster shear rate $\dot{\gamma} \simeq 0.06$~s$^{-1}$.
The quiescent dynamics  show a time dependence similar to that obtained using the other confocal microscope setup (compare figures \ref{msd_shear} and \ref{msd}): particles are localised on the experimentally accessible time window and no long-time diffusion is observed.
One can observe though that the localisation length is larger for the measurements with the shear cell setup.
This might be attributed to the combination of two factors \cite{savin05}: the larger noise level of the multi-beam VT-infinity confocal microscope, which arises from the cross-talk of the fluorescence emission from many different particles simultaneously excited; the smaller magnification (63x instead of 100x) and the larger pixel size (0.25~$\mu$m compared to 0.115~$\mu$m), which increase the uncertainty in the determination of particle coordinates.  
Note also that, in order to compare to measurements under shear, the quiescent MSDs are measured in a two-dimensional plane instead of a three dimensional volume as in the other setup.
Application of a slow shear rate, corresponding to $Pe_{\dot{\gamma}}^l = 6.7\times 10^{-2}$, induces a significant acceleration of the non-affine dynamics of the large particles, as shown in figure \ref{msd_shear}: The particles are initially localised on the same length scale as in the quiescent state but become delocalised at $\gamma\geq 6$~\%, with the MSD increasing first sub-linearly and then linearly with $\gamma$ over the remaining range of measured times. 
The final linear increase of $\delta r_l^2\sin t$ ($\gamma\propto t$) indicates diffusive behavior.
At the larger shear rate ($Pe_{\dot{\gamma}}^l = 4.2\times 10^{-1}$) the particle dynamics first show localisation on a length scale smaller than in the quiescent state and for $\gamma>3$~\% the MSD increases more than linearly with time $t$ and might at larger $\gamma$ tend to normal diffusion. 
For the smaller shear rate, the cage-deformation introduced by shear enables the initially caged particles to diffuse, resulting in the observed acceleration of the average single-particle dynamics and glass melting.
The larger shear rate is sufficiently large to possibly induce cage constriction at short times resulting in the lower localisation length of the MSD. Moreover, the observed super-diffusive behavior could result from the transition from highly constrained in-cage motions to out-of cage shear induced diffusive motions. 
The observed behavior is similar to the one which occurs in one-component colloidal glasses and dense fluids under application of a constant shear rate, as shown in experiments \cite{besseling07,zausch08,besseling09,nick:prl:12,laurati:jpcm:2012}, simulations \cite{varnik06,zausch08,laurati:jpcm:2012} and Mode-Coupling theory \cite{zausch08,laurati:jpcm:2012}. 
In particular, a link between shear-induced cage break up and acceleration of the dynamics has been found \cite{nick:prl:12}. 
Upon application of shear, the cage increasingly deforms, until the maximum elastically sustainable deformation is achieved, where a stress overshoot is observed in rheology, and the cage opens, allowing for diffusion in the steady state of shear where residual structural anisotropy is observed.
Before steady state is achieved, super-diffusion is observed at the transition from caging to diffusion, corresponding to cage yielding.
When the shear rate becomes sufficiently large, cage constriction is continuously induced by shear and a super-diffusive regime is observed in the steady state \cite{nick:prl:12}, similar to what is observed in the mixture for the larger shear rate.  
Note that for $x_s=0.9$ the cage being deformed is that composed of small spheres. 
The $Pe_{\dot{\gamma}}^s=\delta^3 Pe_{\dot{\gamma}}^l$ values for the small spheres are 5.4$\times 10^{-4}$ and 3.4$\times 10^{-3}$ for the slower and faster shear rates respectively. Cage constriction effects are typically observed for $Pe_{\dot{\gamma}} > 0.1$ in one-component glasses \cite{nick:prl:12}.\\
For glasses composed by only one species of particles the long-time diffusion coefficient $D_L$ in the steady state of shear is dominated by the time scale introduced by the shear rate and scales as $\dot{\gamma}^{0.8}$ \cite{besseling07,varnik06}.
The scaling clearly does not hold in this case, since the values on the x-axis scale with $\dot{\gamma}$. The ratio between the $D_L$ values for the two applied shear rates apparently scales with a larger exponent of approximately 1.6, which could be related to the peculiar properties of the mixture.

\section{Conclusions}   
We presented experimental results on the structure, dynamics and viscoelasticity of glasses formed by binary colloidal mixtures with size ratio $\delta = 0.2$ and different mixing ratios. 
Changes in the properties of these glasses as a function of mixing ratio can be rationalized in terms of a transition from caging of the large spheres to caging of the small spheres. 
In comparison to a glass composed of only large spheres, mixing a large fraction of large spheres with a small fraction of small spheres induces pronounced changes in the glass state. 
The cage of large spheres is deformed due to the inclusion of small spheres in the free volume between the large particles. 
This loosening of the cage results in increased mobility of the large particles and an acceleration of their dynamics. 
Correspondingly a strong decrease of the elastic modulus is observed.
Further increasing the fraction of small spheres, the cage distortion increases as more and more small particles fill the free volume. 
This is consistent with random close packing occurring at a larger total volume fraction \cite{biazzo:2009}. 
It also results in a further speeding up of the dynamics and reduction of the elastic modulus. 
At $x_s = 0.5$ on average each large sphere can be completely covered by small spheres and leads to a disruption of the cage structure of the large spheres. 
Concomitantly the response of the system starts to be dominated by caging of the small spheres. 
This is seen as a tighter localisation of the large spheres and a modulus which starts to increase again. 
This trend continues with increasing $x_s$. 
In systems dominated by the cage of the small particles, the large spheres reduce the order on the intermediate length scale beyond the first shell, i.e. the cage.
If shear is imposed on a mixture where caging by the small component dominates the response, the initially frozen dynamics become diffusive in the experimental time-window at small shear rates, and super-diffusive at larger shear rates. A stronger localisation at short times is also observed at larger shear rates.
This indicates that application of shear induces melting of the glass by facilitating out-of-cage diffusion through elongation and deformation of the cage, and cage constriction at large shear rates, similar to recent results on one-component glasses \cite{nick:prl:12,laurati:jpcm:2012}.

\begin{theacknowledgments}
  We acknowledge support from the Deutsche Forschungsgemeinschaft through the FOR1394 Research unit. We also thank G. Petekidis, Th. Voigtmann, K. J. Mutch, P. Chauduri, J. Horbach, M. Fuchs  and W. C. K. Poon for stimulating discussions.
\end{theacknowledgments}



\bibliographystyle{aipproc}   % if natbib is available

\bibliography{laurati_sdcs_v3}

\IfFileExists{\jobname.bbl}{}
 {\typeout{}
  \typeout{******************************************}
  \typeout{** Please run "bibtex \jobname" to optain}
  \typeout{** the bibliography and then re-run LaTeX}
  \typeout{** twice to fix the references!}
  \typeout{******************************************}
  \typeout{}
 }

\end{document}